\documentclass[preprint,aps,prb,floatfix,superscriptaddress]{revtex4-1}
\usepackage{graphics}
\usepackage{subfigure}
\usepackage{epsfig}
\usepackage{verbatim}

\input epsf

\def\be{\begin{equation}}
\def\ee{\end{equation}}
\def\ba{\begin{eqnarray}}
\def\ea{\end{eqnarray}}
\begin{document}

\title{Tunable nano Peltier cooling device from geometric effects using a single graphene nanoribbon}

\author{ Wan-Ju Li}
\affiliation{Department of Physics, Purdue University, West
Lafayette, IN  47907, USA}
\author{Dao-Xin Yao}
\affiliation{State Key Laboratory of Optoelectronic Materials and Technologies,
Sun Yat-sen University, Guangzhou 510275, China}
\author{E.~W.~Carlson}
\affiliation{Department of Physics, Purdue University, West
Lafayette, IN  47907, USA}

%\date{\today}

\begin{abstract}

Based on the phenomenon of curvature-induced doping in graphene we
propose a class of Peltier cooling devices, produced by
geometrical effects, without gating. We show how a graphene nanoribbon
laid on an array of curved nano cylinders can be used to create a
targeted and tunable cooling device. Using two different approaches, the Nonequlibrium Green's Function (NEGF) method and experimental inputs, we predict
that the cooling power of such a device can approach the order of $kW/cm^2$, on par
with the best known techniques using standard superlattice structures. The
structure proposed here helps pave the way toward designing graphene electronics which use geometry rather than gating to control devices.

\end{abstract}

\maketitle

As electronics become smaller, one problem facing the industry is that of
targeted, on-demand cooling, conventionally achieved by
semiconductor-based Peltier coolers\cite{zeng2004,chowdhury2009,fan2001,shakouri2005,zhang2003}, containing arrays of n-type and p-type pellets. However, one limitation
of Peltier coolers made with conventional fabrication
techniques is that carrier concentration for each pellet is set by doping level and cannot be adjusted after fabrication, which in turn results in that the cooler's cooling power cannot be dynamically changed. Therefore, new materials and novel structures are desired to fabricate a cooler with dynamically changeable cooling power.

Graphene, a two-dimensional atomic layer, is a potential candidate for a high-performance thermoelectric cooler. As a result of its unique energy spectrum, the doping level can be controlled by shifting the Fermi level either above (n-type) or below (p-type) the Dirac point. Therefore, remarkable efforts have been focused on both electrical\cite{chiu2010,liu2008,nam2011,oezyilmaz2007,rao2011,williams2007,yu2011} and chemical\cite{cheng2011,lohmann2009} methods for fabricating graphene-based p-n junctions, which are the main constituents of a Peltier cooler. However, due to the geometry of these p-n junctions, the direction of  heat transfer is parallel to the fabrication substrate.  For this reason, the geometry of coolers has to be such that the p- and n-type regions are perpendicular to the surface which needs to be cooled, like a stack of books on a shelf. This extra level of manufacturing makes the devices less likely to be of use, especially in nanoscale applications.

\begin{figure}
\includegraphics[width=.95\columnwidth]{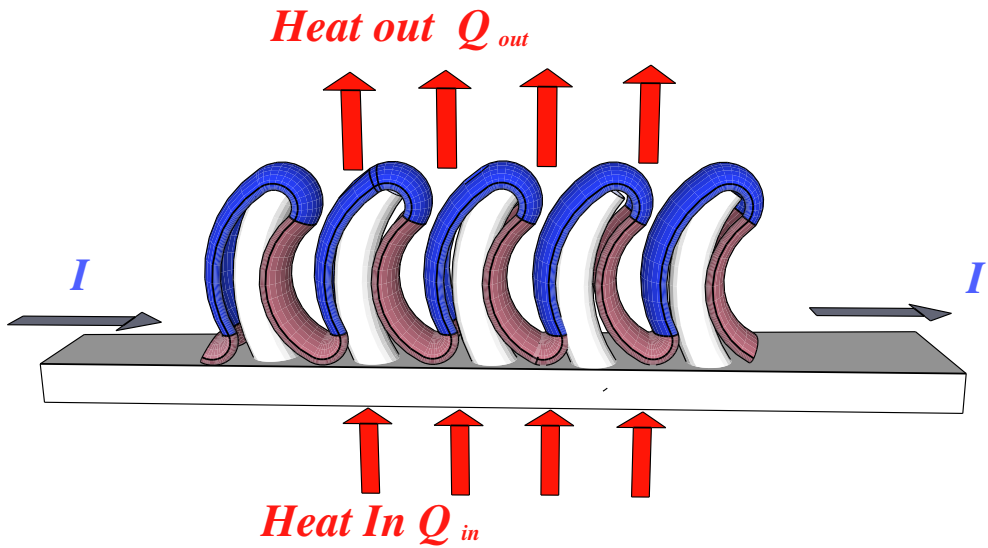}
  \caption{A single GNR laid on an array of curved nanotubes. A GNR bent into the geometry shown
becomes a Peltier cooling device with cooling power
$\approx$1kW/cm$^2$. The large mean curvature  of the blue regions
leads to spontaneous n-type doping, and the low mean curvature of
the red regions leads to spontaneous p-type doping. Upon application
of a current from left to right, heat is pumped from bottom to top.
The cooling power may be adjusted by changing the curvature,
via, e.g., the application of uniaxial pressure on the device.}
  \label{3d1}
\end{figure}

Whereas the currently available fabrication techniques of the graphene-based p-n junction are confined to surfaces, the {\em three dimensional} configurations of this flexible 2D membrane may eventually be exploited to develop integrated
circuits with components not achievable before. In this paper,
we show how a single, continuous graphene
nanoribbon (GNR) may be used to create a nanoscale Peltier array without
need for lithography or gating.  The key ingredients in any Peltier cooler are an
array of p-n junctions which are electrically in series yet thermally in parallel, such that
the junctions which evolve heat are on one side of the device, and
the junctions which absorb heat are on the other. Upon application
of a current from left to right through the GNR shown in Fig. \ref{3d1},
heat will be pumped from the junctions near the bottom of the
structure to the junctions near the top. From two different methods which agree, we estimate the
cooling power of the proposed device to be on the order of $kW/cm^2$.

\begin{figure}
\centering
\subfigure[two curvature radii]{
\includegraphics[width=.45\columnwidth]{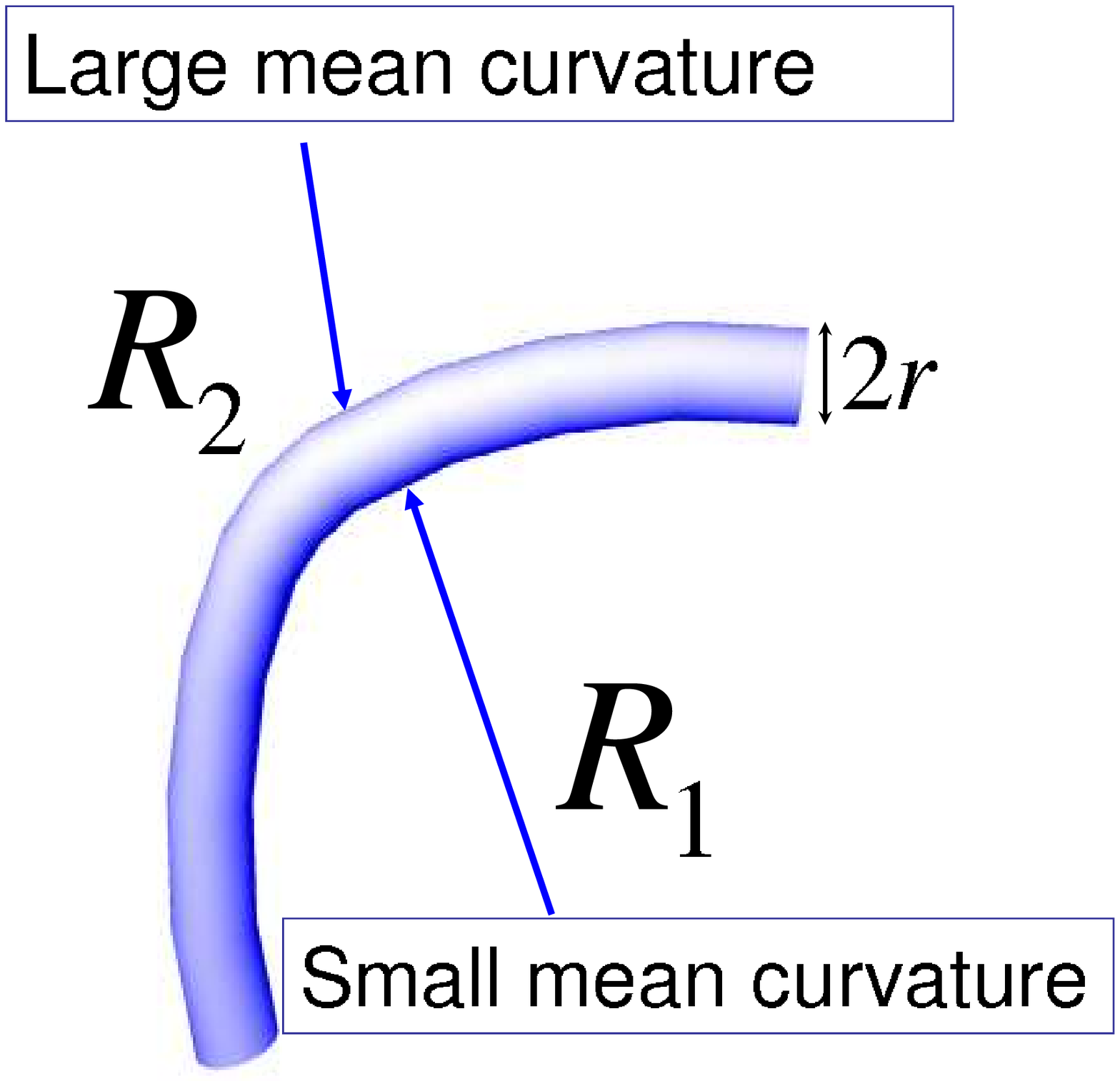}
\label{fig:twocurvature} } \subfigure[Procedures for building the proposed device.]{
\includegraphics[width=.45\columnwidth]{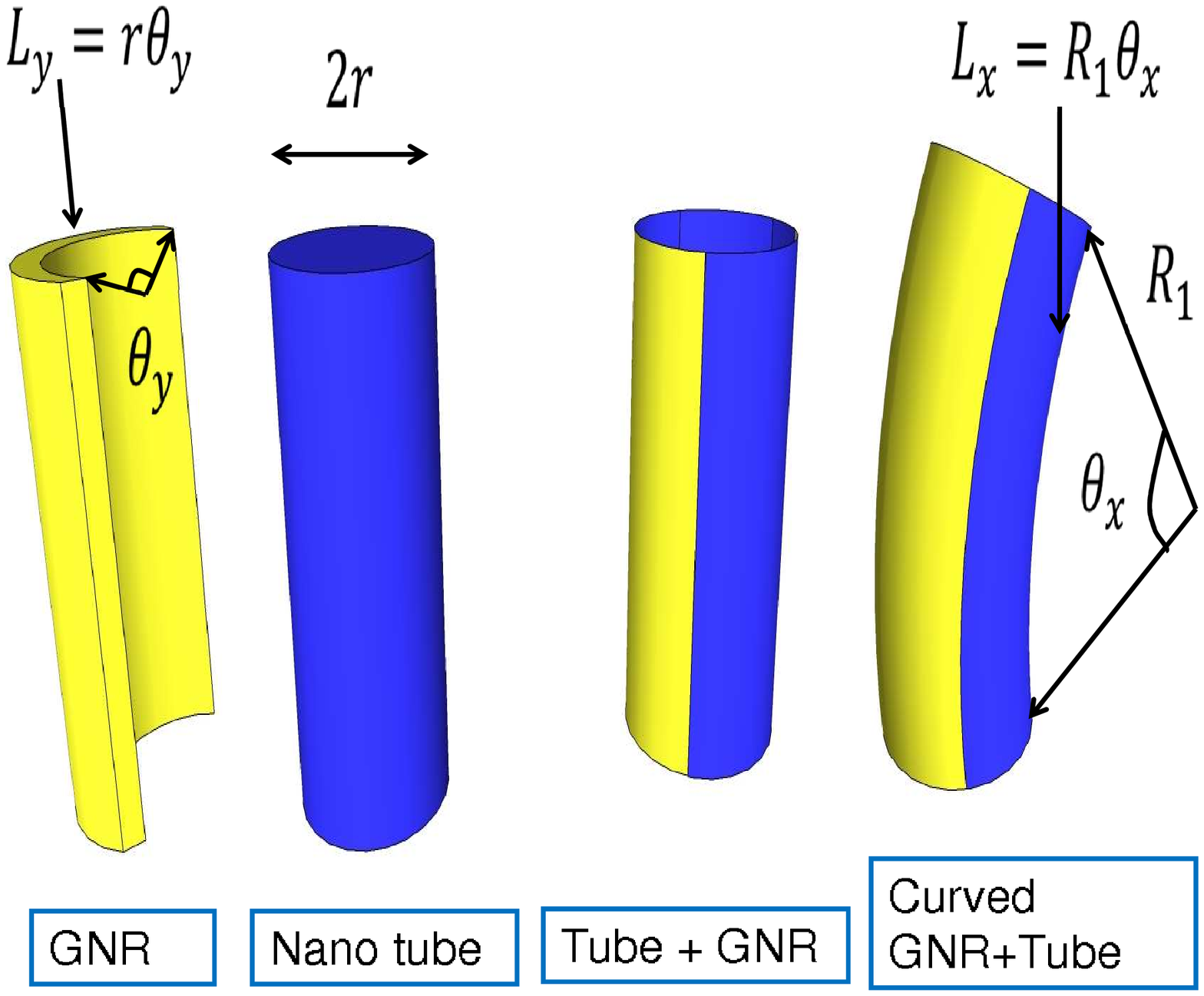}
\label{fig:thetaxy} }
\caption{(a). Creating two curvatures by bending. Two curvature radii with equal signs results in large mean curvature while two radii with different signs results in small mean curvature.
$R_1$ and r are the two radii for the inner side and $R_2$ and $r$ are for the outer side of the tube. (b). A schematic illustration of making our proposed nanocooler. At first, a GNR is put on the surface of a insulating nanotube. Secondly, curvatures are introduced on the GNR by bending the nanotube. The curving angle $\theta_x$ ($\theta_y$) defined by $L_x$ and $R_1$ ($L_y$ and $r$).} \label{fig:proposal}
\end{figure}

Curvature-induced doping provides a route to creating the p-n
junctions required by Peltier coolers.
When the local curvature of graphene
resembles that of a sphere (large mean curvature), or that of a saddle
(small mean curvature), the change in orbital energies and overlap
integrals associated with the curvature causes the energy spectrums and the Dirac points
to shift differently in these two cases, leading to n-type
or p-type doping\cite{kim2008}. By draping a single Armchair metallic GNR over a curved cylindrical protrusion (such as a bent
nanotube may provide, see Fig. \ref{fig:proposal}\subref{fig:thetaxy}), a region of large mean curvature is created next to a region of small mean curvature, and a p-n junction is created. 

In Fig. \ref{3d1},
blue regions denote areas of large mean curvature and red regions denote
areas of small mean curvature. The curvature-induced Dirac-point
shift is given by\cite{kim2008} 
\begin{equation}
\Phi(\mathbf{x})=-3 \alpha a^2 \bigg(H(\mathbf{x}) \bigg)^2~.
\end{equation}
$H(\mathbf{x})$ is the mean curvature of the surface at the
point $\mathbf{x}$
\begin{equation}
H =\frac{1}{2} \bigg( {1 \over r_1} + {1 \over r_2}\bigg)~,
\end{equation}
where $r_1$ and $r_2$  are the principal radii of curvature at the
point $\mathbf{x}$, $a$ is the nearest-neighbor distance, $a=2.5\AA$, and $\alpha\sim 9.23eV$. In our proposed device, see Fig. \ref{fig:proposal}\subref{fig:twocurvature}, $r_1=R_2$ and $r_2=r$ for the outer surface while $r_1=R_1$ and $r_2=r$ for the inner surface.  Because the Dirac-point shift depends on the
square of $H$\cite{kim2008}, it is independent of the coordinate system. The Dirac points in the regions
with a larger magnitude of the mean curvature (locally like a
sphere) are lowered more while the Dirac points in the regions with smaller
magnitude of the mean curvature (regions which are flat, or
alternatively which are saddle-like) are lowered less. This effect implies that regions with large magnitude of the mean curvature are locally electron-doped, while the 
regions with small magnitude of the mean curvature are locally hole-doped.

The optimal cooling power $P_m$ for one Peltier cooling element can be derived from the rate of heat absorption\cite{rowe2006b}
\begin{equation}
P_m=K(\frac{1}{2}ZT_C^2-\Delta T)=\frac{(\Delta S)^2
    T_C^2}{2R}-K\Delta T,
    \label{cooling}
\end{equation}
where $\Delta T=T_H-T_C$ is the temperature difference between the two heat reservoirs, $\Delta S=S_p-S_n$, is the difference between Seebeck coefficients in the p-doped (with smaller mean curvature) and n-doped (with larger mean curvature) regions, and $Z=\frac{(\Delta S)^2}{KR}$, the figure of merit of the p-n junction. $K$ and $R$ are the thermal conductance and electric resistance, respectively, of the p-n junction. When electrical current equal to $ I=I_{opt}=\frac{(\Delta S) T_C}{R}$ is driven through one Peltier cooling element, the optimal cooling power $P_m$ is reached. 

In the following, we use two approaches to estimate the cooling power of our proposed device under the condition that $\Delta T=T_H-T_C=0$. For the first approach, we use Nonequilibrium Green's Function (NEGF) method to perform calculations from the atomic scale. For the second approach, we estimate the cooling power from experimental measurements of the Seebeck coefficient\cite{wei2009}. Results from these two approaches are similar. 

%NEGF start
The NEGF method can be used to calculate the transmission coefficient $T_0(E)$ for a perfect GNR from the atomic scale\cite{datta2005b}. The system contains one channel and two contacts. The Green's function $G$ for the transport channel is defined as:

\begin{equation}
G=[(E+i\eta)I-H-\Sigma_1-\Sigma_2]^{-1},
\end{equation}

where $E$ is the incident energy of electrons, $I$ is the identity matrix with the same dimension as that of the channel Hamiltonian $H$, $\eta$ is a small number accounting for the energy level broadening, and $\Sigma_1$ and $\Sigma_2$ represent the effects from two contacts. In order to describe the tunnelling processes between the channel and two contacts, two functions, $\Gamma_1$ and $\Gamma_2$, are introduced for two contacts respectively,

\begin{equation}
\Gamma_1=i[\Sigma_1-\Sigma_1^+]; \Gamma_2=i[\Sigma_2-\Sigma_2^+].
\end{equation} 
The transmission coefficient for a perfect GNR is then given as\cite{datta2005b}

\begin{equation}
T_0(E)=Trace[\Gamma_1 G \Gamma_2 G^+].
\label{transmission}
\end{equation}

In real systems, scattering processes have to be taken into account. The electron-electron interaction usually causes phase relaxation, which destroys the coherence of the electronic wave functions. The electron-phonon interaction and the electron-impurity interaction result in backscattering and momentum relaxation processes. For high voltage bias between two contacts, effects from inelastic scatterings are also involved. Dominant scattering processes are determined by the physical system under consideration. 

Under usual experimental conditions, where the voltage bias is small and the whole system is in the linear response regime, there is a way to take elastic scattering effects into account within NEGF. For quantum transport, it is well known that the transmission coefficient $T(E)$ can be related to its ballistic limit counterpart $T_0(E)$ in the following way\cite{datta2005b},  

\begin{equation}
T(E)=\frac{\lambda}{\lambda +L}T_0(E),
\end{equation} 
where $L$ is the system size along the transport direction and $\lambda$ is the backscattering mean-free path and is assumed to depend only on disorder and not on the incident energy $E$. Therefore, once $T_0(E)$ is obtained, it is straightforward to calculate $T(E)$. For this paper, calculations of the the transmission coefficient are based on the GNR system with $L_x=75nm$, $L_y=25nm$, and $\lambda=400nm$. 

\begin{figure}
\centering
\subfigure[The electric conductance for perfect GNR.]{
\includegraphics[width=.45\columnwidth]{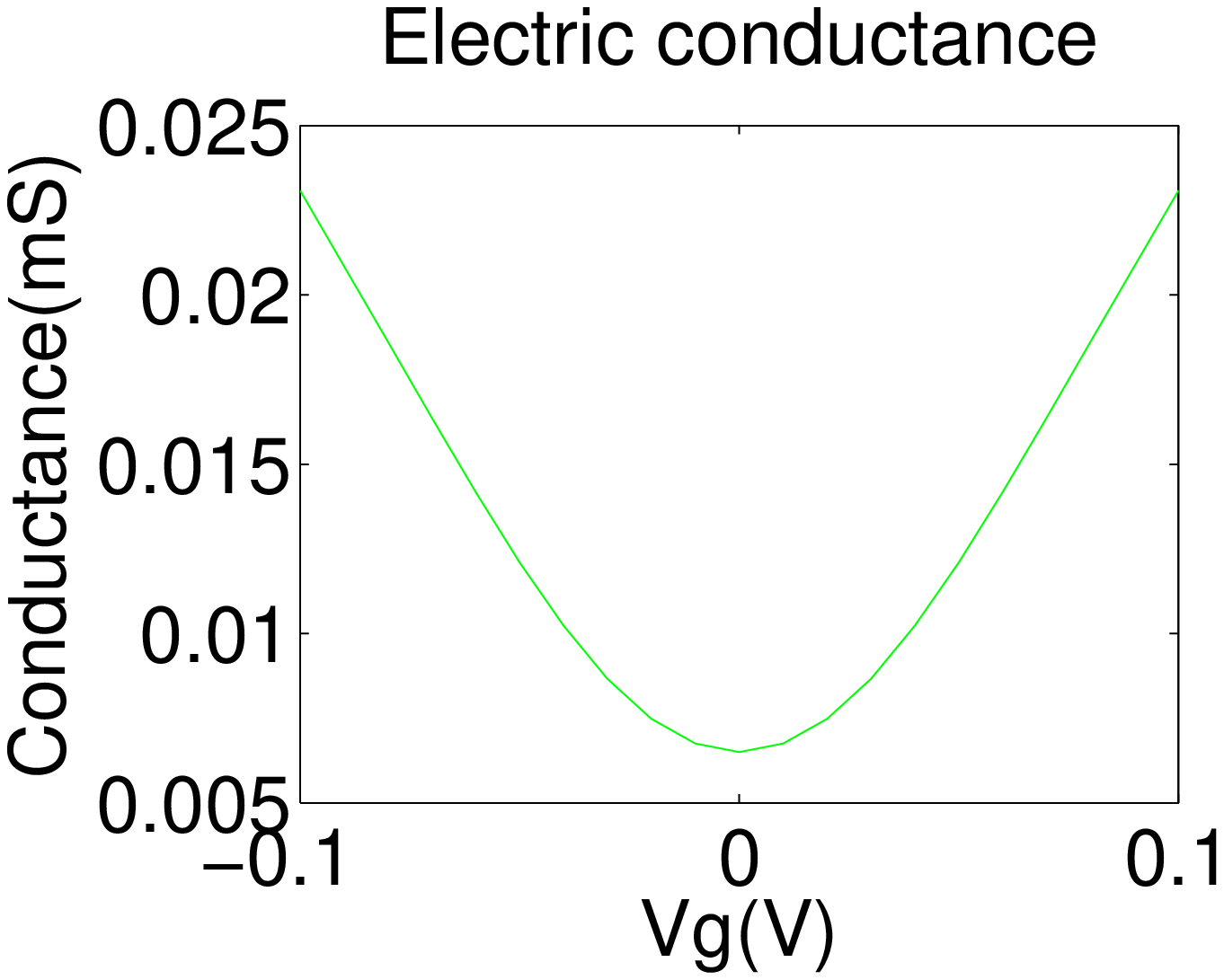}
\label{fig:ecclean}} 
\subfigure[The electric conductance for disordered GNR.]{
\includegraphics[width=.45\columnwidth]{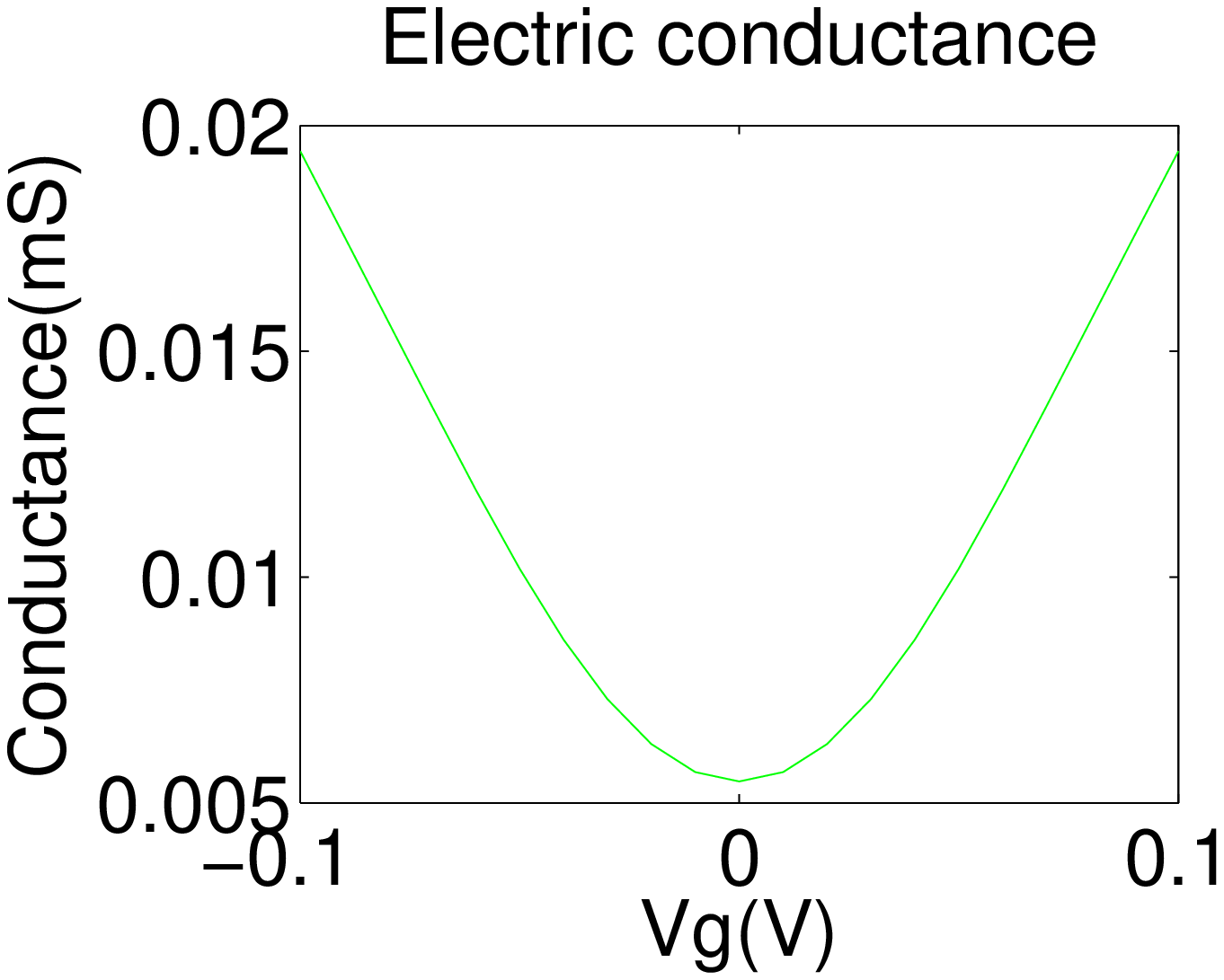}
\label{fig:ecdisorder} }
\subfigure[The Seebeck coefficient for perfect GNR.]{
\includegraphics[width=.45\columnwidth]{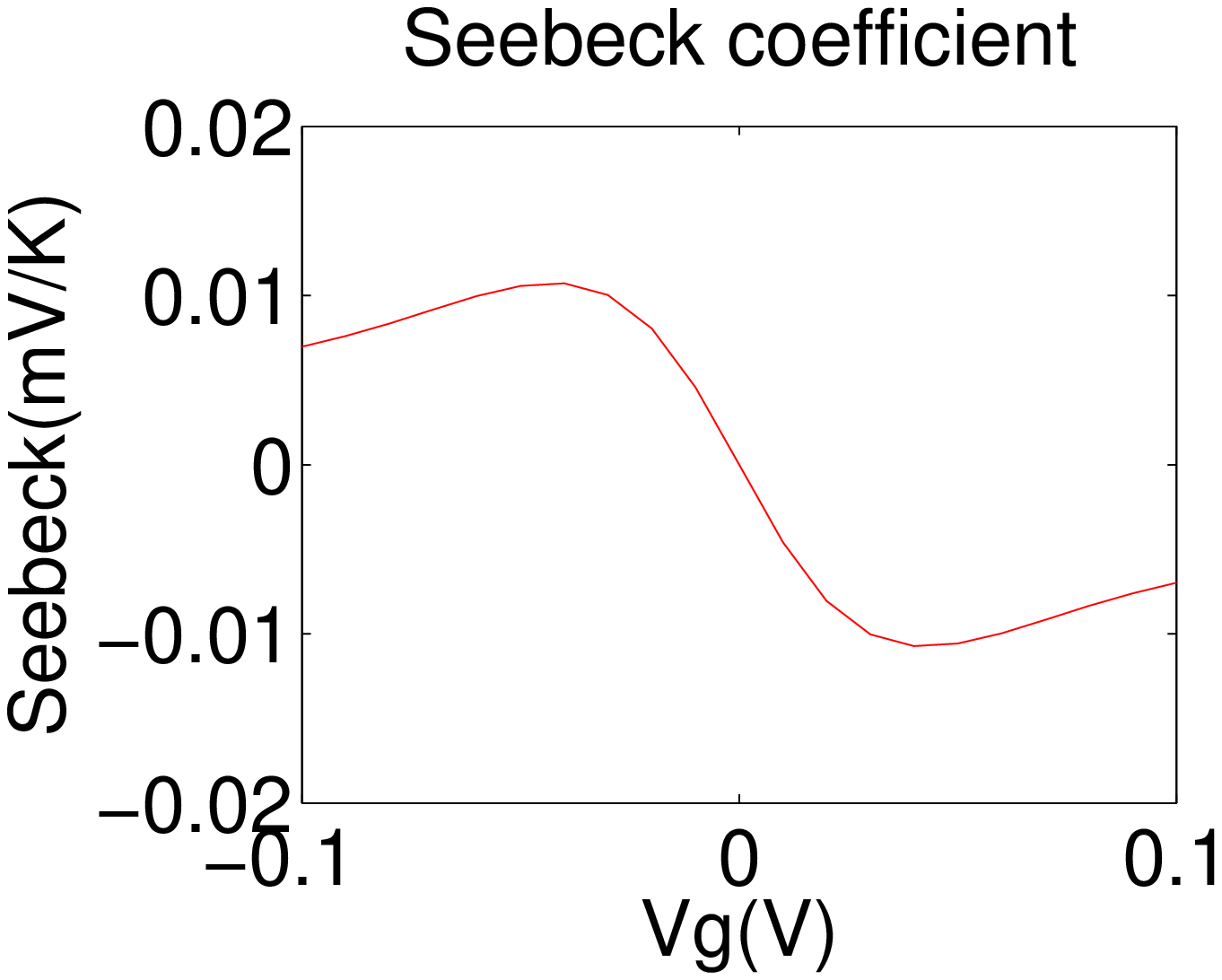}
\label{fig:seeclean} } 
\subfigure[The Seebeck coefficient for disordered GNR.]{
\includegraphics[width=.45\columnwidth]{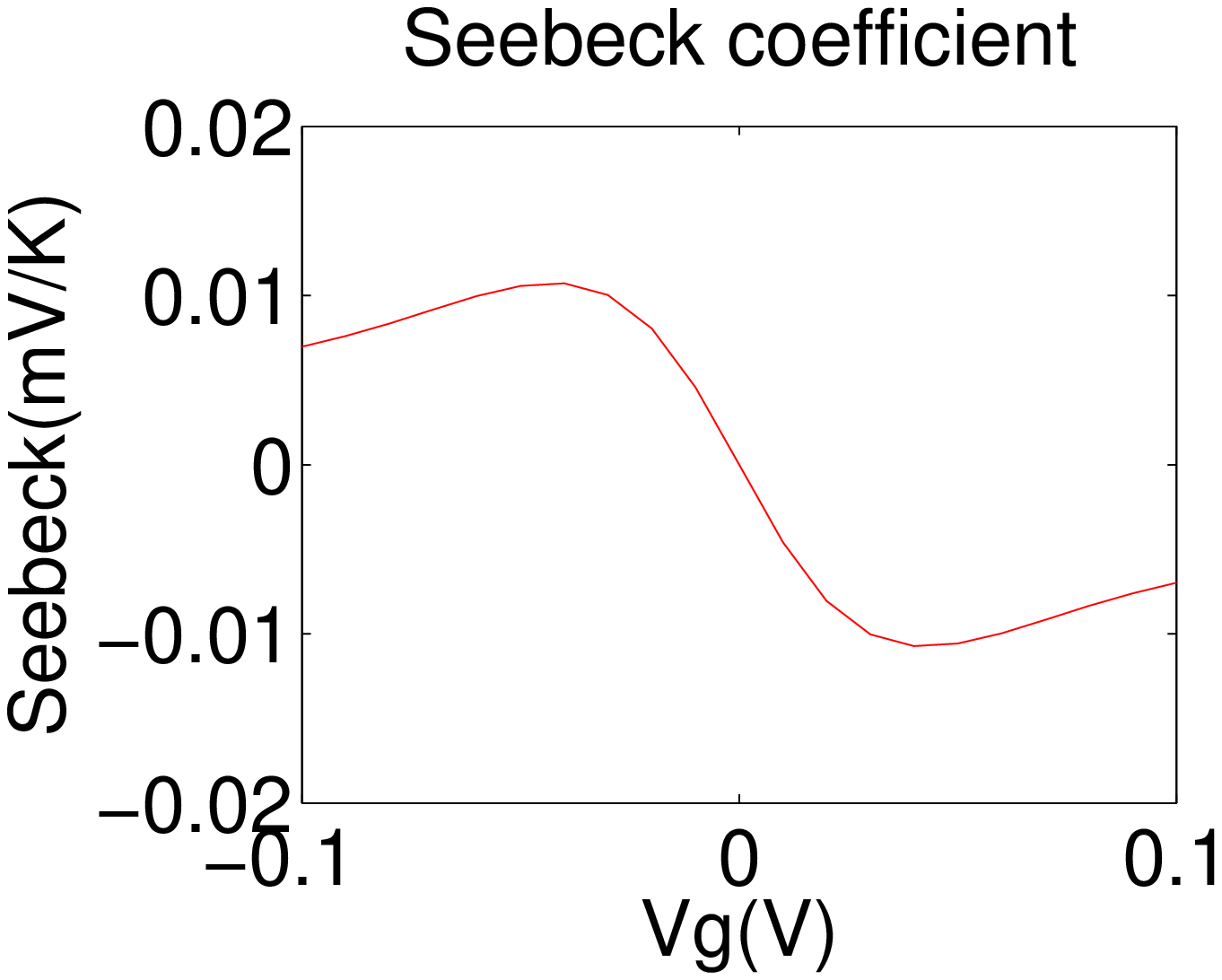}
\label{fig:seedisorder} }
\caption{The sample size is that $L_x=75nm$, $L_y=25nm$ and $T=300K$.(a). The electric conductance for the perfect GNR(b). The electric conductance for the disordered GNR with $\lambda = 400nm$.(c). The Seebeck coefficient as a function of gate voltage for the perfect GNR.(d). The Seebeck coefficient as a function of gate voltage for the disordered GNR with $\lambda = 400nm$.} \label{fig:negf}
\end{figure}

After obtaining $T(E)$ we can calculate the Seebeck coefficient $S$ and the electrical conductance $g$. An intermediate function $L_n$ is defined as\cite{ouyang2009}

\begin{equation}
L_n(\mu,T)=\frac{2}{h}\int dE T(E)(E-\mu)^n(-\frac{\partial f(E,\mu,T)}{\partial E}),
\end{equation}
where $h$ is the Plank constant, and $f$ is the Fermi distribution function. Based on these intermediate functions, $g$ and $S$ can be computed as

\begin{equation}
g(\mu)=e^2 L_0(\mu,T),
\end{equation}

\begin{equation}
S(\mu)=\frac{1}{eT}\frac{L_1(\mu,T)}{L_0(\mu,T)},
\end{equation}
where $e$ is the electron charge, $\mu$ is the chemical potential and $T$ is the average temperature. In Fig.\ref{fig:negf}, we show the electric
conductance and the Seebeck coefficient at $T=300K$ as functions of the on-site voltage for clean and disorder cases. We find that the electrical conductance of the clean case is slightly larger than that of the disordered case while the Seebeck coefficient behaves similarly in both cases. We furthermore find that the maximum of the Seebeck coefficient is on the order of $100 \mu V/K$, consistent with the experimental measurements for Graphene\cite{wei2009}. 

%It implies that, at room temperature, electrons in a GNR with width $L_y=25nm$ behave similarly to those in a graphene sheet in micrometers.    

The cooling power $P$ of our proposed device can be evaluated using the calculated Seebeck coefficient and electrical conductance with disorder. When the surface of a GNR is curved, Dirac points
for different regions are shifted up or down, similar to applying a gate voltage.
Therefore, the calculated Seebeck coefficient as a function of on-site
voltage can be used as the Seebeck coefficients
for different Dirac-point shifts caused by the curvatures. Combined with the calculated electric conductance, the cooling power can be calculated by using Eq.\ref{cooling} with $\Delta T=0$. 

\begin{figure}
\includegraphics[width=.95\columnwidth]{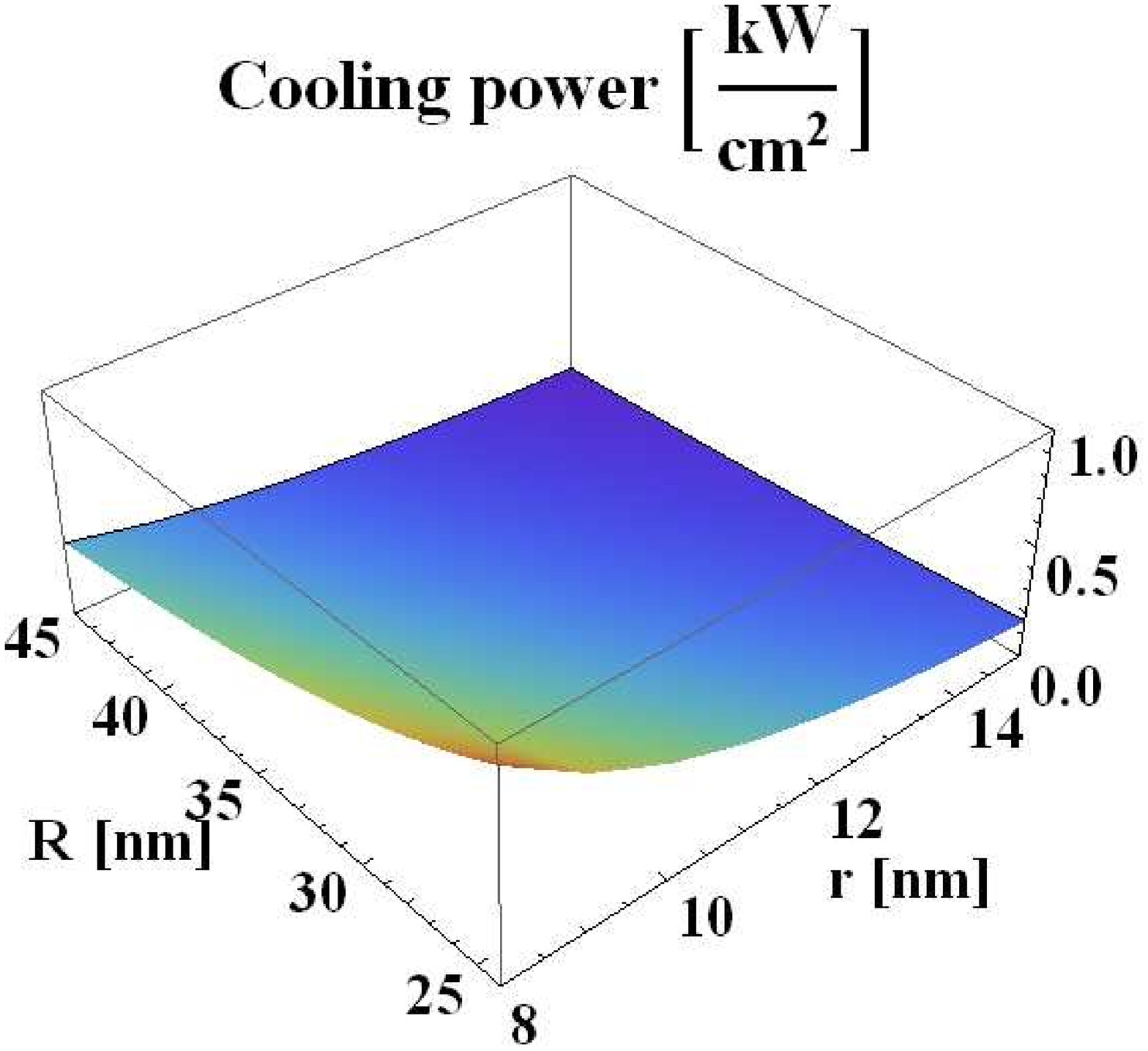}
  \caption{The cooling power as a function of two curvature radii for the GNR with $L_x=75nm$, $L_y=25nm$, and $\lambda = 400nm$, which is from NEGF calculations.}
  \label{coolnegf}
\end{figure}

In Fig. \ref{coolnegf}, we show the cooling power as a function of two curvature radii $R=R_1$ and $r$, where $R=\frac{L_x}{\theta_x}$, $r=\frac{L_y}{\theta_y}$, $T=300K$, and the distance between two nano tubes $d=15nm$. $L_x$ and $L_y$ are fixed and $0<\theta_x,\theta_y<\pi$. Therefore $R$ and $r$ have lower limits $\frac{L_x}{\pi}$ and $\frac{L_y}{\pi}$, respectively. As we expect, the maximum cooling power takes place for small $R$ and $r$, which is on the order of $kW/cm^2$. As an example, if we take as inputs $L_x=75nm=\theta_x R$, $L_y=25nm=\pi r$, and $\theta_x=\frac{\pi}{2}$, then the corresponding Dirac-point shifts for the inner side ($\Phi_1$) and
outer side ($\Phi_2$) can be obtained after considering charge neutrality : $\Phi_1\approx 1.96$ $mV$, and $\Phi_2\approx-1.96$ $mV$. By using the calculated Seebeck coefficient and the electrical conductance (Fig. \ref{fig:negf}\subref{fig:seedisorder} and Fig. \ref{fig:negf}\subref{fig:ecdisorder}, respectively), we find that the cooling power is $~0.3kW/cm^2$. When the cooling device is curved more, $\theta_x=\frac{2\pi}{3}$, the cooling power can be estimated in a similar way to be $~0.5kW/cm^2$. This shows that the cooling power can be tuned by changing the curvature of the device, for example, by applying uniaxial pressure.

%NEGF end
We now use an entirely different approach to estimate the cooling power of our proposed device, in which
the Seebeck coefficients are taken directly from experimental measurements(Fig. 3(b) in ).
 Given the applied gate voltage $V_g$, the resulting Dirac-point
shift of the Graphene sample $V_r$ can be derived as\cite{novoselov2004} 
\begin{equation}
V_r=\frac{\hbar v_F}{e}\sqrt{\frac{\epsilon_0\epsilon \pi}{te}}\sqrt{V_g},
\end{equation}
where $t$ is the thickness of the $S_i O_2$ substrate, $v_F$ is the Fermi velocity of graphene, and $\epsilon_0$ and $\epsilon$ are the permittivities of free space and $S_i O_2$, respectively. By using experimentally determined Seebeck coefficient\cite{wei2009} and electrical conductivity\cite{novoselov2004}, the cooling power can be similarly estimated from Eq. \ref{cooling}.  

\begin{figure}
\includegraphics[width=.95\columnwidth]{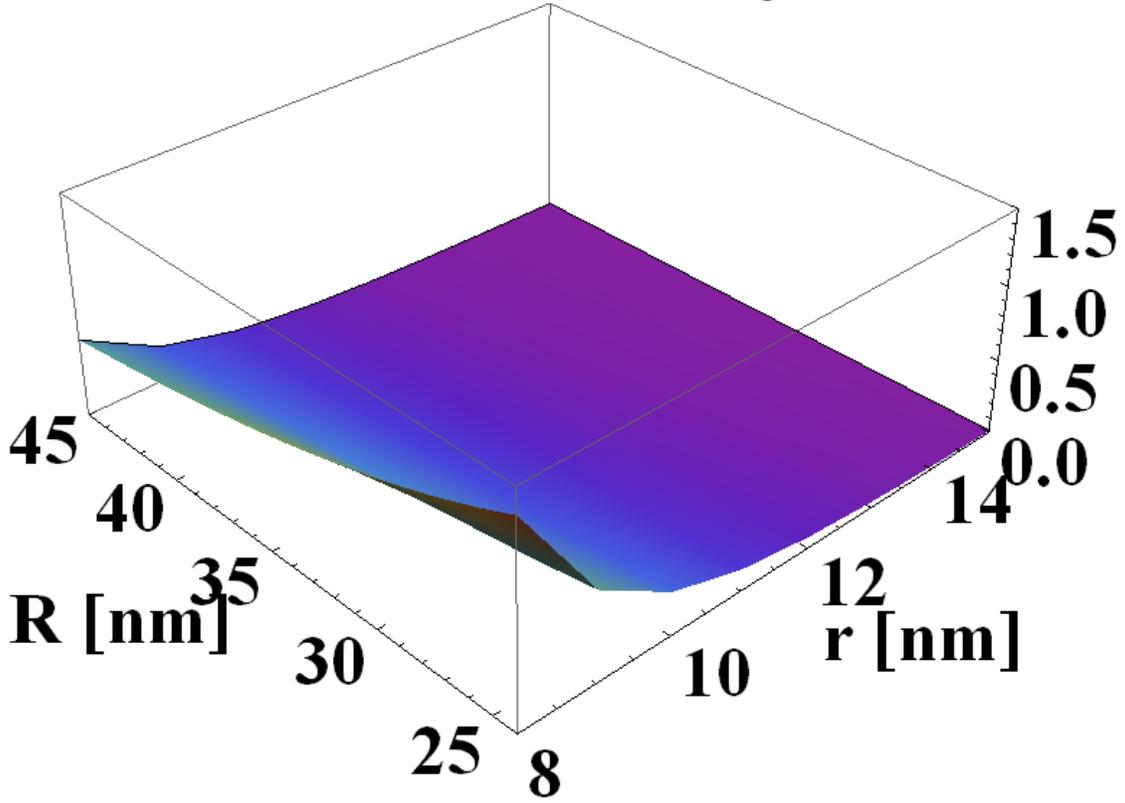}
\caption{The cooling power as a function of two curvature radii, $R$ and $r$ for a GNR with $L_x=75nm$, $L_y=25nm$. The calculation is performed by using the experimental input\cite{wei2009}}
\label{coolpower}
\end{figure}

In Fig. \ref{coolpower}, we report the cooling power as functions of the two curvature radii $r$ and $R$ as we do for the first approach.  Results are consistent with that from the first approach, which is on the order of $kW/cm^2$. As an example, if we take as inputs $L_x=75nm=\theta_x R$, $L_y=25nm=\pi r$, and $\theta_x=\frac{\pi}{2}$, then the corresponding Dirac-point shifts for the inner side ($\Phi_1$) and
outer side ($\Phi_2$) can be obtained: $\Phi_1\approx 1.96$ $mV$,
and $\Phi_2\approx-1.96$ $mV$. This corresponds to applied gate voltages
of $\Phi_{1g}\approx 1.38$ $V$, and $\Phi_{2g}\approx -1.38$
$V$. Given these two gate voltages, the Seebeck coefficients in both the n-type and p-type regions\cite{wei2009} can be obtained. Combined with experimental data for the electric conductance(~$10^6 \frac{1}{\Omega - m}$, corresponding to the mobility ~$10^4 \frac{cm^2}{V-S}$), the cooling power $P$ can be estimated to be $P\approx 0.57 \frac{kW}{cm^2}$. When the cooling device is curved more, $\theta_x=\frac{2\pi}{3}$, from similar calculations, the cooling power is obtained to be $\approx 0.9
\frac{kW}{cm^2}$. This result again shows that we can tune the cooling power by
bending the nanotubes through, e.g., applying uniaxial pressure.

In conclusion, we have proposed a graphene-based nano mechanical
cooling device and estimated its cooling power using two different approaches: the NEGF method and experimental inputs. As a result of geometry alone, a series of P-N junctions are created in the proposed
device such that by applying electric current, heat can be pumped perpendicular to the surface of the substrate. We find $P\sim 0.5\frac{kW}{cm^2}$, close to that achievable with the best cooling devices $\sim 1\frac{kW}{cm^2}$\cite{chowdhury2009}. Most importantly, the cooling power of the proposed device can be adjusted by changing the curvatures, via, e.g., applying uniaxial pressure to the device.

{\bf Acknowledgements}~~~It is a pleasure to thank Y. Chen, E.-A. Kim,
and Y. L. Loh  for conversations. W. J. L. would like to thank Vinh Quang Diep and Seokmin Hong for many useful discussions.
W. J. L., D. X. Y., and E. W. C. acknowledge support from Research Corporation
for Science Advancement and NSF Grant No. DMR 11-06187
W. J. L. acknowledges support from the Purdue Research Foundation. D. X. Y. acknowledges support from NSFC-11074310, National Basic
Research Program of China (No. 2012CB821400), and Research Fund for the Doctoral Program of
Higher Education of China (20110171110026). 
EWC thanks \'Ecole Sup\'erieure de Physique et de Chimie Industrielles
(ESPCI) for hospitality.

%\bibliographystyle{phjcp}
%\bibliography{graphenepaper}

\end{document}